\documentclass[12pt]{iopart}

\usepackage[svgnames,table]{xcolor}
\usepackage[T1]{fontenc}
\usepackage{txfonts}
\usepackage{graphicx}

\usepackage{hyperref}
\hypersetup{
        colorlinks=true,
        citecolor=SteelBlue,
        filecolor=LimeGreen,
        linkcolor=SlateBlue,
        urlcolor=MediumPurple
}

\makeatletter
\let\cat@comma@active\@empty
\makeatother
\begin{document}
%\preprint{APS/123-QED}
\title{A six degree-of-freedom fused silica seismometer: Design and tests of a metal prototype}
\author{Amit Singh Ubhi, Jiri Smetana, Teng Zhang, Sam Cooper, Leo Prokhorov, John Bryant, David Hoyland, Haixing Miao, and Denis Martynov}
\address{Institute for Gravitational Wave Astronomy, School of Physics and Astronomy, University of Birmingham, Birmingham B15 2TT, United Kingdom}

\date{\today}

\begin{abstract}
Ground vibrations couple to the longitudinal and angular motion of the aLIGO test masses and limit the observatory sensitivity below 30\,Hz. Novel inertial sensors have the potential to improve the aLIGO sensitivity in this band and simplify the lock acquisition of the detectors. In this paper, we experimentally study a compact 6D seismometer that consists of a mass suspended by a single wire. The position of the mass is interferometrically read out relative to the platform that supports the seismometer. We present the experimental results, discuss limitations of our metallic prototype, and show that a compact 6D seismometer made out of fused silica and suspended with a fused silica fibre has the potential to improve the aLIGO low frequency noise.

\end{abstract}

\maketitle

\section{Introduction}\label{sec:intro}

    The network of the Advanced LIGO~\cite{Aasi2015} and Advanced Virgo~\cite{Acernese2014} gravitational-wave (GW) detectors observed tens of signals from compact objects with masses up to $150\,M_\odot$~\cite{Abbott2016,Collaboration2016,Collaboration2017,Collaboration2017a,Abbott2020}. Most of the astrophysical sources were categorised as neutron stars and stellar mass black holes, with one event producing an intermediate-mass black hole~\cite{Abbott2020}. The signals were mostly accumulated at frequencies from 30\,Hz up to 500\,Hz due to the diminished response of the detectors at lower and higher frequencies. Despite the sophistication and successful operation of aLIGO’s seismic isolation platforms~\cite{Matichard_2015,MATICHARD2015287}, the sensitivity degrades below 30\,Hz due to non-stationary control noise~\cite{Yu2018,Buikema2020, Martynov_Noise_2016}.
    
    Improvements of the aLIGO inertial isolation will result in a number of benefits (i-v) for the detectors.
    %(i) routine observations of intermediate-mass black-holes, (ii) improved localisation of the sources and enable early-warning alerts to search for the electromagnetic counterparts, (iii) simplification of the lock acquisition process, (iv) enhanced duty cycle, and even (v) improvement of the detector sensitivity in their most sensitive frequency band around 100\,Hz. 
    Enhanced sensitivity below 30\,Hz will lead to (i) routine observations of intermediate-mass black-holes since heavier compact objects merge at lower frequencies. Observing GWs at lower frequencies also increases the measurement time of mergers~\cite{Yu2018} and provides (ii) an improved localisation of the sources and enables early-warning alerts to search for the electromagnetic counterparts of events like GW170817~\cite{Collaboration2017a}.
    %The combination of these improvements increases the likelihood of precise localisation of sources.
    Furthermore, since the aLIGO lock acquisition process is a complex procedure that involves stabilisation of five longitudinal and tens of angular degrees of freedom~\cite{Staley_LOCK_2014, Martynov_THESIS_2015}, suppression of the test mass motion will (iii) simplify the lock acquisition process. The coincident duty cycle of the aLIGO Livingston and Hanford detectors has the potential to (iv) increase from the current 62.2\%~\cite{Buikema2020} up to $\approx 75$\% once seismic distortions are further suppressed. Finally, reduced motion of the aLIGO test masses will allow us to increase the beam size in the arm cavities and (v) improve the coating thermal noise~\cite{Harry_Thermal_2012, Gras_CTN_2017} that limits the aLIGO sensitivity in its most sensitive frequency band around 100\,Hz~\cite{Buikema2020, Martynov_QUCORR_2017}.
    
    %The key problem with larger beam sizes comes from the angular motion of the test masses that make the interferometer less stable. Improving the low frequency platform motion will also reduce the overall motion of the test masses and make the interferometer stable.
    
    %which relinquishes the strain on the angular control loops which are only used to resonate stable power within the arms. Without reducing the bandwidth of these control loops, the differential arm (DARM) signal receives injection of the control loop noise. Ideally, we would like to lower the bandwidth of angular controls from current 3 Hz to ~0.1 Hz.\note{CITATIONS NEEDED}
    
    %These noises are caused 
    
    A number of inertial sensors has been proposed in the literature~\cite{Venkateswara2014,MowLowry2019, Collette2015, Korth_Gyro_15, Martynov_Gyro_2019, Heijningen_2018, McCann_2021} that have the potential to reduce the aLIGO platform motion and enhance the detector sensitivity. The key problem comes from the coupling of the aLIGO platform tilt to its horizontal motion. This coupling scales very unfavourably as $g/\omega^2$, where $g$ is the local gravitational acceleration, and $\omega$ is the angular frequency of the platform motion~\cite{Lantz-tilt,Lantz2009,Matichard2015,Matichard2016}. The aLIGO detectors already utilise custom beam rotation sensors~\cite{Venkateswara2014} developed by the University of Washington to measure the tilt motion of the ground. The next logical step would be to develop new custom sensors for the aLIGO platforms to further suppress their motion.

    In this paper, we experimentally investigate a compact 6D seismometer that has the potential to measure the aLIGO platform motion in all six degrees of freedom~\cite{MowLowry2019}.
    The 6D seismometer is similar to drag-free control, which is a technique employed in satellites to keep the spacecraft a constant distance from a free-falling mass~\cite{Theil2008}.
    In contrast to the rotation sensors, we avoid mechanical constraints of the inertial mass and allow it to move freely in all degrees of freedom.
    The mechanical simplicity, however, leads to digital complexity since we need to diagonalise coupled degrees of freedom.
    In Sec.~\ref{sec:prototype}, we discuss experimental results, show how to diagonalise the coupled degrees of freedom, and compare the sensitivity of the 6D seismometer with commercial sensors. In Sec.~\ref{sec:design}, we propose a vacuum-compatible version of the compact 6D seismometer and discuss the impact for the aLIGO detectors.
    
    % The low tilt modes enables this decoupling of the tilt to translational motion to enhance the low frequency sensitivity of the detectors. This upgrade to aLIGO will increase the probability of detecting IMBH and make the detector more robust.
    
    % Increase the beam size results in a reduction in coating thermal noise as the incident power flux on the mirror is reduced.
    % Large beam size makes controls more difficult, better/more robust control allows for increase in beam size to improve coating noise in most sensitive frequency band.
    % Larger beam sizes improve coating noise but also increases the arm cavity g-factor. In other words, angular motion of the mirrors lead to higher power fluctuations in the interferometer if beam sizes are increased. If we make the beam size larger then all these noises will be higher unless we suppress the platform motion with compact 6D or similar sensors. We only need angular controls to resonate stable levels of power in the interferometer. The drawback of angular controls is that they leak to the longitudinal degree of freedom and corrupt DARM. Ideally, we would like to lower the bandwidth of angular controls from current 3 Hz to ~0.1 Hz or so. But for this we need to suppress the LIGO platform motion. If we are very successful, we can even potentially increase the beam size in the arms.
    
\section{Metallic prototype}\label{sec:prototype}
    
\begin{figure}[t]
        \centering
        \includegraphics[width = 0.9\columnwidth]{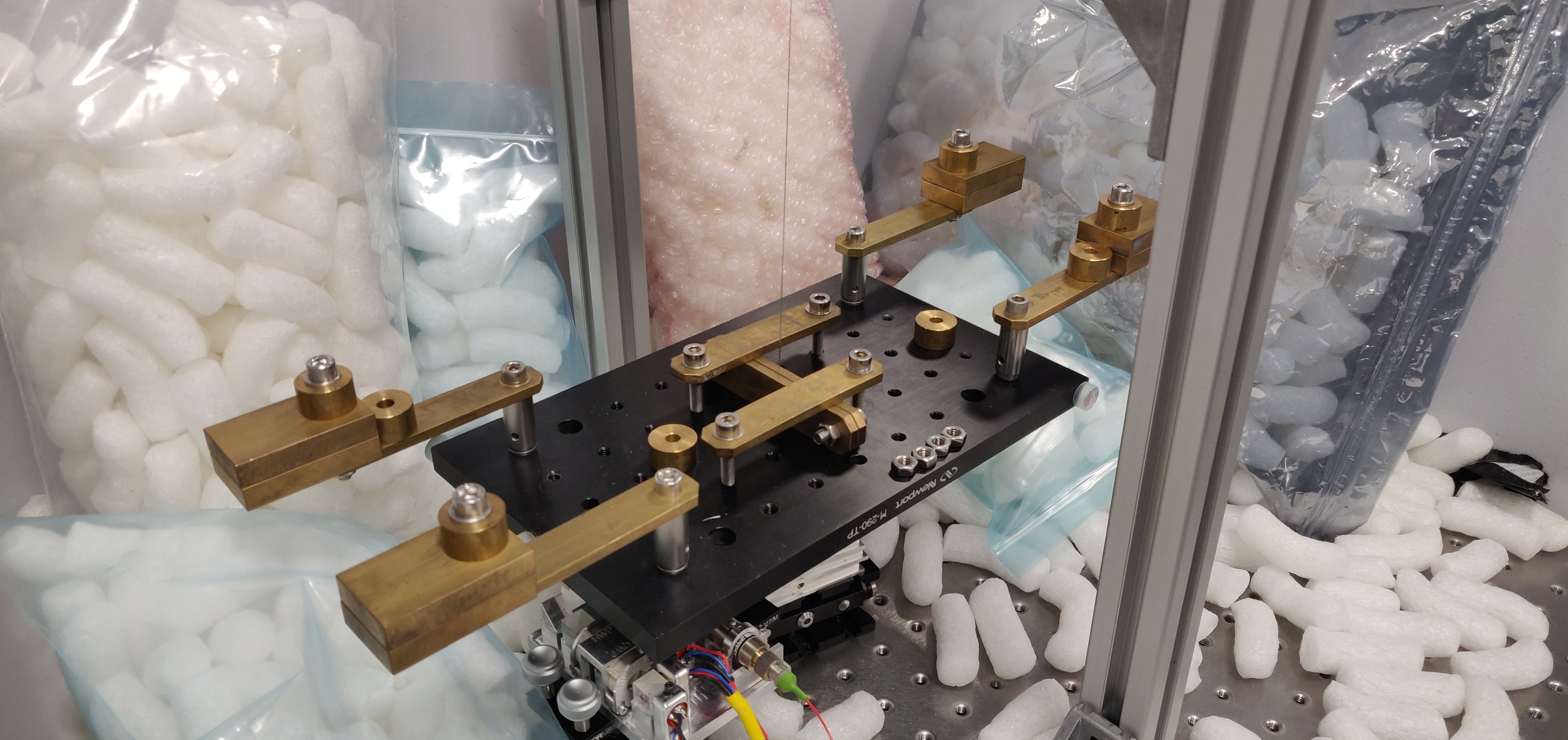}
        \caption{Picture of the isolated metallic prototype with an interferometric sensor.}
        \label{fig:prototype}
    \end{figure}

    In this section, we discuss a metallic prototype, shown in Fig.~\ref{fig:prototype}, that was constructed with the aim of understanding the mechanics and dynamics of the system. The purpose was to provide insight for the fused silica design discussed in Sec.~\ref{sec:design}.
    
\subsection{Mechanical design}\label{sec:opto-mechanical-design}
    
    The prototype used a 21\,cm long, $r = 200 \,\mu \rm m$ radius stainless steel wire to suspend an aluminium breadboard of dimensions ($254 \times 152.4 \times 12.7$) mm - this was used as the base for the proof mass. We added additional brass masses to extend the proof mass and increase its moment of inertia to $I_{\rm RZ} \approx 0.076\, \rm kgm^2$, $I_{\rm RY} \approx 0.061\, \rm kgm^2$, $I_{\rm RX} \approx 0.015\, \rm kgm^2$, where RX, RY, and RZ implies rotation around the horizontal X- and Y-axes, and the vertical Z-axis. With these parameters, we achieved longitudinal eigenmodes along the X- and Y-axes of $1.096\,\rm Hz$. We set the vertical bounce mode of the suspension at 25\,Hz. The rigidity of the vertical mode made the system impermeable to vertical motion of the ground below 10\,Hz and simplified the measurement of the angular motion of the inertial mass. This can be seen in Fig.~\ref{fig:spectra}, where above 10\,Hz the vertical commercial L4C seismometer \cite{L4Cwebsite} (orange) measures the same as the vertical sensor on the proof mass (red), and below this, the vertical motion is not dominant in the sensor. The torsional (RZ) eigenmode was set at 20\,mHz.
    
    The tilt (RX and RY) eigenmodes of the suspended mass depend on the vertical centre-of-mass position of the system according to the equation
    \begin{equation}
        \centering
        \omega_{\rm RX,RY}^2 = (2\pi f_{\rm RX,RY})^2 \approx \frac{m g \Delta z + k_{\rm el}}{I_{\rm RX,RY}},
        \label{eq:tilt-resonance}
    \end{equation}
    where $\Delta z$ is the centre of mass position relative to the bending point of the bottom part of the wire, $k_{\rm el} = 1/2 \sqrt{m g E I_a}$ is the elastic restoring coefficient of the wire, $m=3.6$\,kg is the mass of the seismometer mass, $E=210$\,GPa is the Young's modulus of the steel wire, and $I_a = \pi r^4 /4$ is the second area moment of inertia of the wire.

    For the configuration used, the tilt modes were $f_{\rm RY} = 84\,\rm mHz$ and $f_{\rm RX} = 167\,\rm mHz$ respectively. We adjusted $\Delta z$ to create a gravitational anti-spring and lower the tilt modes. The centre of mass position was above the bending point by $\approx 0.8$\,mm, creating an inverted pendulum. The resulting response was such that the effective spring constant $k_{\rm eff} = m g \Delta z + k_{el}$ was smaller than the restoring coefficient of the wire, $k_{\rm el}$. We ensured that the effective spring constant still had a positive value so that the system remained stable. The key problem with reducing $k_{\rm eff}$ was that balancing the mass became difficult due to the plastic deformation of the steel wire as discussed in Sec.~\ref{sec:plastic}.
    %We achieved the precision of the vertical centre of mass tuning, $\Delta z$, equal $\sim$ few $\mu \rm{m}$.
    
    The system had quality factors of 2000 for the horizontal modes and 700 for the tilt modes. These quality factors were limited by air damping, and by mechanical losses in the metal fibres and in the clamping mechanisms. The large quality factors for a steel wire suspension led to significant motion of the inertial mass. We reduced the quality factors down to 70 and 3.5 for the longitudinal and tilt modes with eddy current damping. The effect of damping can be seen in Fig.~\ref{fig:transfer-functions} as the resonant peaks are broad.
    
    %Another effort to combat the large motion below 100\,mHz was to use eddy current damping, which also allowed for ease in alignment of the system as the drift motion was reduced. The effect of this can be seen in Fig.~\ref{fig:transfer-functions} as the resonant features are broad. Without damping the system had Q factors of 2000 for the translational mode and 700 for the pitch mode compared to 70 and 3.5 respectively. However temperature fluctuations in the lab can cause these high Q resonances to change marginally, making system identification more challenging resulting in difficulty in decoupling.
    
    % The prototype focused on maximising the moment of inertia around the y-axis (pitch), $I_{ry}$ (see Fig.\,\ref{fig:prototype}) in order to decrease the eigenmode.
    
    %Eqn.\,\ref{eq:tilt-resonance} shows the relationship between the tilt resonance $\omega_0$ and the various parameters which effect it. There is also the capability of adjusting the centre of mass position relative to the bending point $\Delta z$ such that it has negligible effect compared to the elastic spring restoring constant $k_{el} = 1/2 \sqrt{m g E I_a}$, where $m$ is the mass of the proof mass, $E$ is the Youngs modulus of the fibre, and $I_a = \pi r^4 /4$ is the area moment of inertia with $r$ being the wire radius;
    
\subsection{Plastic deformations}
\label{sec:plastic}

The key experimental challenge of balancing the softly suspended mass came from plastic deformations in the steel wire. The wire stress due to the wire tension is given by the equation
\begin{equation}
    \sigma_t = \frac{m g}{\pi r^2} = 281\,{\rm MPa}.
\end{equation}
The stress, $\sigma_t$, was already close to the yield stress of the wire material (350\,MPa) during the experiment. In addition to the stress from the tension, $\sigma_t$, DC tilts of the mass led to the extra stress near the surface of the wire given by the equation
\begin{equation}
    \sigma_{\rm RY} \approx k_{\rm el} \frac{\alpha_{\rm RY}}{r^3} = 500 \frac{\alpha_{\rm RY}}{0.1\, \rm rad}\,{\rm MPa},
\end{equation}
where $\alpha_{\rm RY}$ is the DC misalignment of the suspended mass around the RY axis. The total stress inside the wire has exceeded the yield stress for misalignment angles of only $\approx 10^{-2}$\,rad. As a result of the plastic deformation, the equilibrium position of the suspended mass continuously changed during the balancing procedure that consisted of adding small masses to either side of the mass. We found that plastic deformations made the balancing procedure ineffective when $k_{\rm eff}$ approached zero. 
    
\subsection{Optical readout}\label{sec:optical-readout}

\begin{figure}[t]
        \centering
        \includegraphics[width = 0.9\columnwidth]{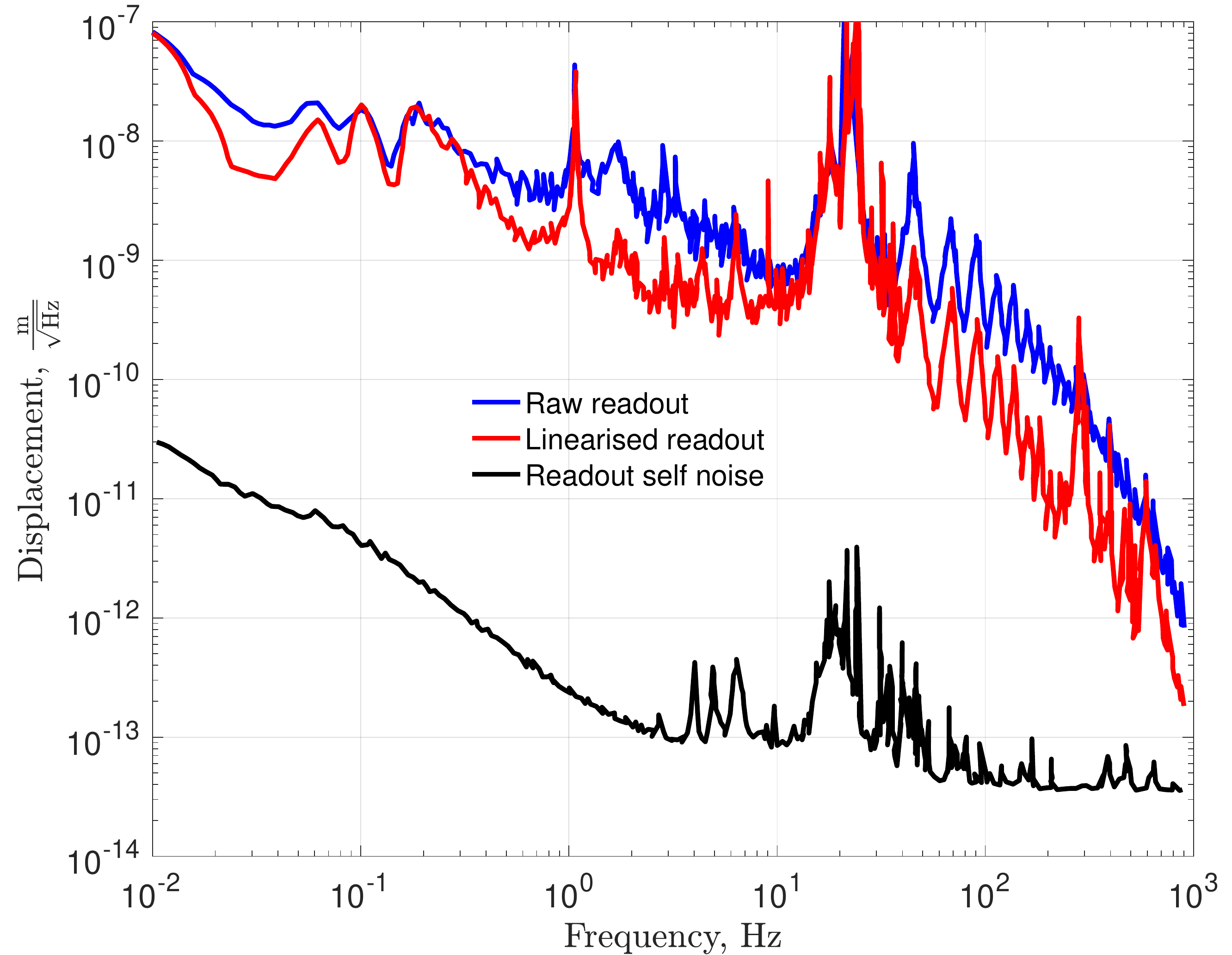}
        \caption{Example of nonlinearities witnessed in the interferometric readout during the experiment. The nonlinearities are strongly suppressed by fitting the Lissajous figures of the compact interferometers with ellipses. The black curve shows the self noise of the sensor measured with stationary mirrors.}
        \label{fig:hoqi}
    \end{figure}

    Measurement of the proof mass was accomplished using interferometric readout with two Homodyne Quadrature Interferometers (HoQIs)~\cite{Cooper_2018}. We utilised only two sensors since this number is sufficient to understand the key cross-couplings between X and RY, and between Y and RX degrees of freedom as discussed in Sec.~\ref{sec:mech-cross-coupling}. One horizontal HoQI measured the translational motion (X), and another HoQI measured the vertical position at the end of the proof mass. The vertical motion was directly converted to RY motion of the mass below 10\,Hz since the vertical mode of the suspension was rigid as discussed in Sec.~\ref{sec:opto-mechanical-design}.
    %the vertical bounce mode was kept rigid, this resulted in the sensor being insensitive to vertical motion of the platform below a 10 Hz, as the signal would then be dominated by the pitch motion of the system.

    HoQIs are compact Michelson interferometers that measure the position of the test mass relative to the reference mirror. The longitudinal range of the sensors is $>10$\,cm since HoQIs measure the position with two polarisations and count optical fringes of the Lissajous figures~\cite{Cooper_2018}. The angular working range of the sensors is 1\,mrad, therefore, isolation from environmental disturbances was necessary to keep the test mass within the HoQI working range as discussed in Sec.~\ref{sec:sensitivity}. Ideally, in order to achieve the best performance of the HoQI sensors, the RMS motion of the mass should be less than one fringe of the interferometer to reduce nonlinear effects in the data. However, the linearisation algorithms~\cite{Cooper_2018,Watchi2018} improve the broadband noise and up-conversion of large motion in the spectra as shown in Fig.~\ref{fig:hoqi}. The linearisation can be further improved above 30\,Hz by a more accurate ellipse fitting of the Lissajous figures~\cite{Cooper_2018,Watchi2018}.
    %The linearisation could be improved via reduction of the RMS displacement of the proof mass, however we are concerned with low frequency motion below 25\,Hz which does not significantly contribute to the RMS motion.
    
\subsection{Mechanical cross-couplings}\label{sec:mech-cross-coupling}

\begin{figure}[t]
        \centering
        \includegraphics[width = 0.9\columnwidth]{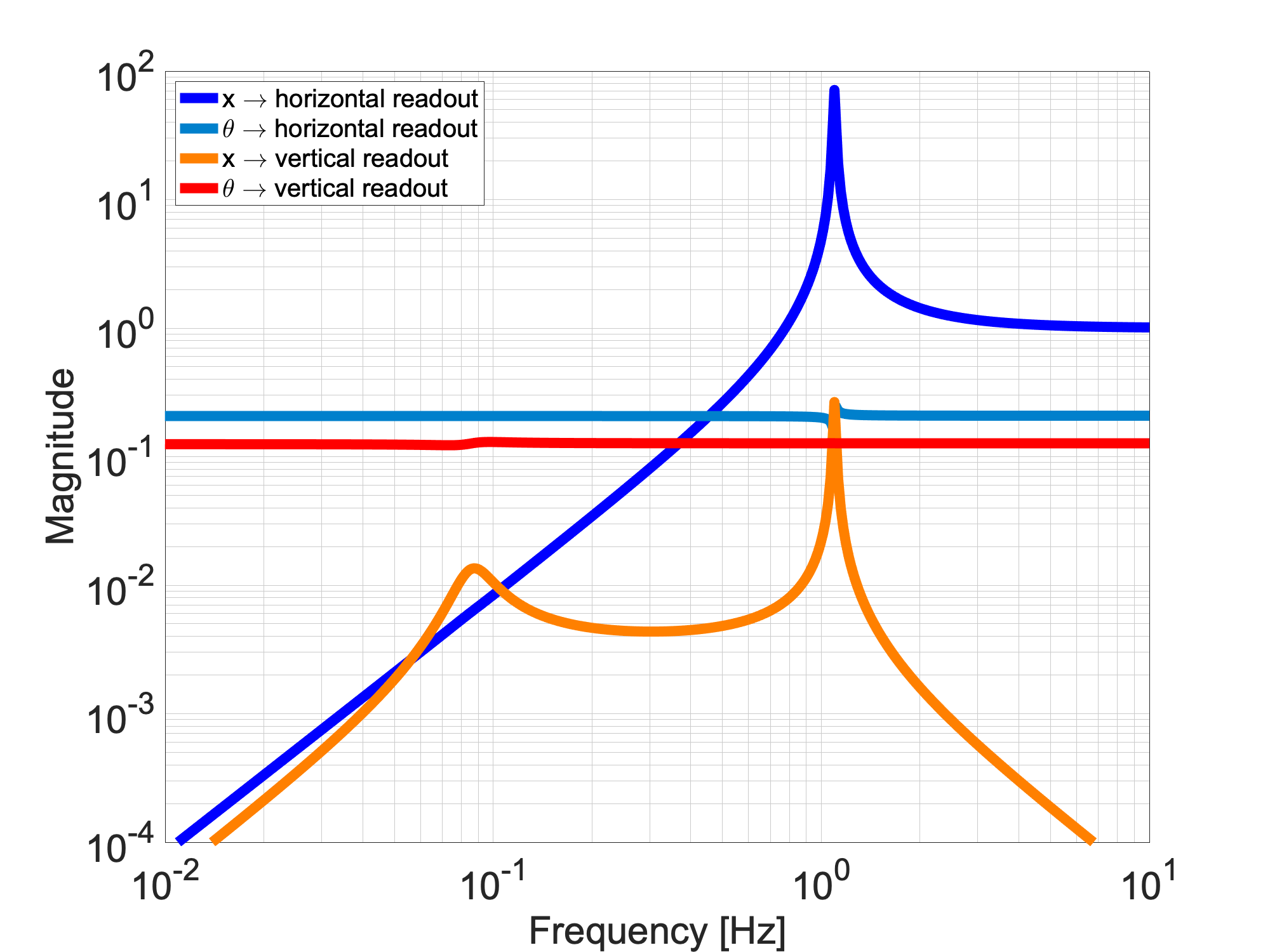}
        \caption{Transfer functions illustrating the coupling between the horizontal and vertical readouts due to the input platform translation $x$, and platform tilt $\theta$.}
        \label{fig:transfer-functions}
    \end{figure}

    In the analytical study of the system dynamics~\cite{MowLowry2019}, we found that the vertical (Z) and torsion (RZ) modes are independent from the other degrees of freedom. For a well aligned and balanced system, the cross-couplings only occur for the X and RY degrees of freedom and, similarly, for the Y and RX degrees of freedom. Further cross couplings can occur due to sensor misalignment, which will result in a more complex transfer function matrix. In this paper, we neglect second order cross couplings and experimentally investigate X and RY frequency-dependent coupling and its diagonalisation.
    
   Fig~\ref{fig:transfer-functions} shows the couplings between the platform motion in X and RY degrees of freedom and readout of the inertial mass position. The platform motion in X strongly couples to the horizontal readout at and above the suspension resonance in X. Below the resonance, the inertial mass moves in common with the platform and the seismometer response is suppressed. In contrast to the 1D tiltmeters~\cite{Venkateswara2014}, the platform X motion also couples to the RY motion of the mass. Tilt motion of the platform has a minor effect on the mass motion in X and RY but moves the local sensors relative to the inertial mass and strongly couples to the readout.
   
   In the case where we infer the X motion of the platform by looking only at the horizontal readout, we will get the typical tilt-to-horizontal coupling~\cite{Lantz-tilt,Lantz2009,Matichard2015,Matichard2016}. This coupling is present in all 1D seismometers and is given by the equation
    \begin{equation}
        \centering
        \hat{x} = x - \frac{g}{\omega^2}\theta,
        \label{eq:tilt-coupling}
    \end{equation}
    where $x$ is the translational platform motion, $\theta$ is its RY motion, and $\hat{x}$ is the inferred motion of the platform. In order to avoid the coupling given by Eq.(\ref{eq:tilt-coupling}), we infer the horizontal motion of the platform from both the horizontal and angular positions of the test mass according to the equation
    \begin{equation}\label{eq:decoupling}
        \begin{pmatrix}
        \hat{x} \\
        \hat{\theta}
    \end{pmatrix}
    =
    \begin{pmatrix}
        T_{\rm x \rightarrow h} & T_{\theta \rm \rightarrow h} \\
        T_{\rm x \rightarrow v} & T_{\theta \rm \rightarrow v}
    \end{pmatrix}^{-1}
    \begin{pmatrix}
        x_m \\
        z_m
    \end{pmatrix},
    \end{equation}
    where $x_m$ and $z_m$ are horizontal and vertical measurements of the HoQI sensors as discussed in Sec.~\ref{sec:optical-readout} and the frequency dependent transfer matrix $T$ is analytically derived and shown in Fig.~\ref{fig:transfer-functions}.
    
    % For the metal prototype, an eigenmode of 84\,mHz was achieved via the use of a stainless steel wire with a radius of $200 \,\mu \rm{m}$. We find that the thinner wire results in a smaller elastic restoring force further reducing the resonance, however practical constraints limit this, as the wire must support the entire suspended mass without plastic deformation such that the system behaves as expected. 
    
    %Initially a metallic prototype design has been constructed which is more compact than the final design discussed in section\,\ref{sec:design}. The aim was to understand the mechanics and behaviours of the system, providing insight into the complexities involved. Achieving low resonant modes for the tilt degrees of freedom is paramount to enable decoupling of the translational and tilt modes, which makes the tilt modes more orthogonal \note{Transfer function figure to show translation to tilt coupling}. An improved tilt signal at 10mHz - 1Hz results in improvement in the tilt decoupling from the translation degree of freedom which effects DARM. \note{citations needed}

\subsection{Sensitivity}\label{sec:sensitivity}

    \begin{figure}[t]
        \centering
        \includegraphics[width = 0.9\columnwidth]{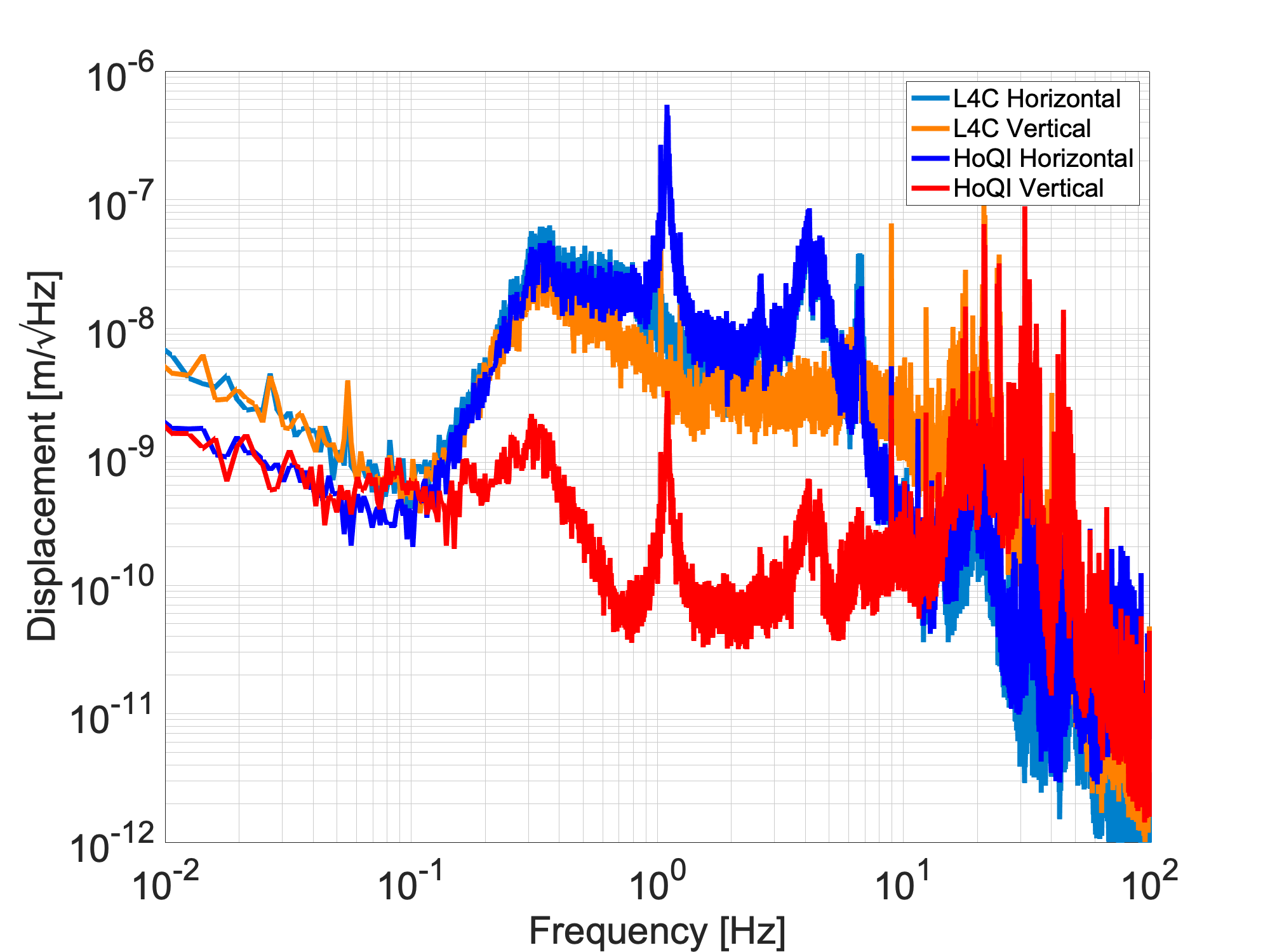}
        \caption{Comparison of measured motion between the compact 6D seismometer and commercial L4C geophones. The translational readout of the seismometer (blue) is similar to the L4C signal (light blue) above 100\,mHz. The vertical readout (red) measures tilt motion of the inertial mass up to 10\,Hz and vertical motion of the platform above this frequency. The vertical L4C (orange) measures the same motion as the vertical HoQI above the bounce mode resonance at 25\,Hz.}
        \label{fig:spectra}
    \end{figure}
    
    \begin{figure}[t]
        \centering
        \includegraphics[width = 0.9\columnwidth]{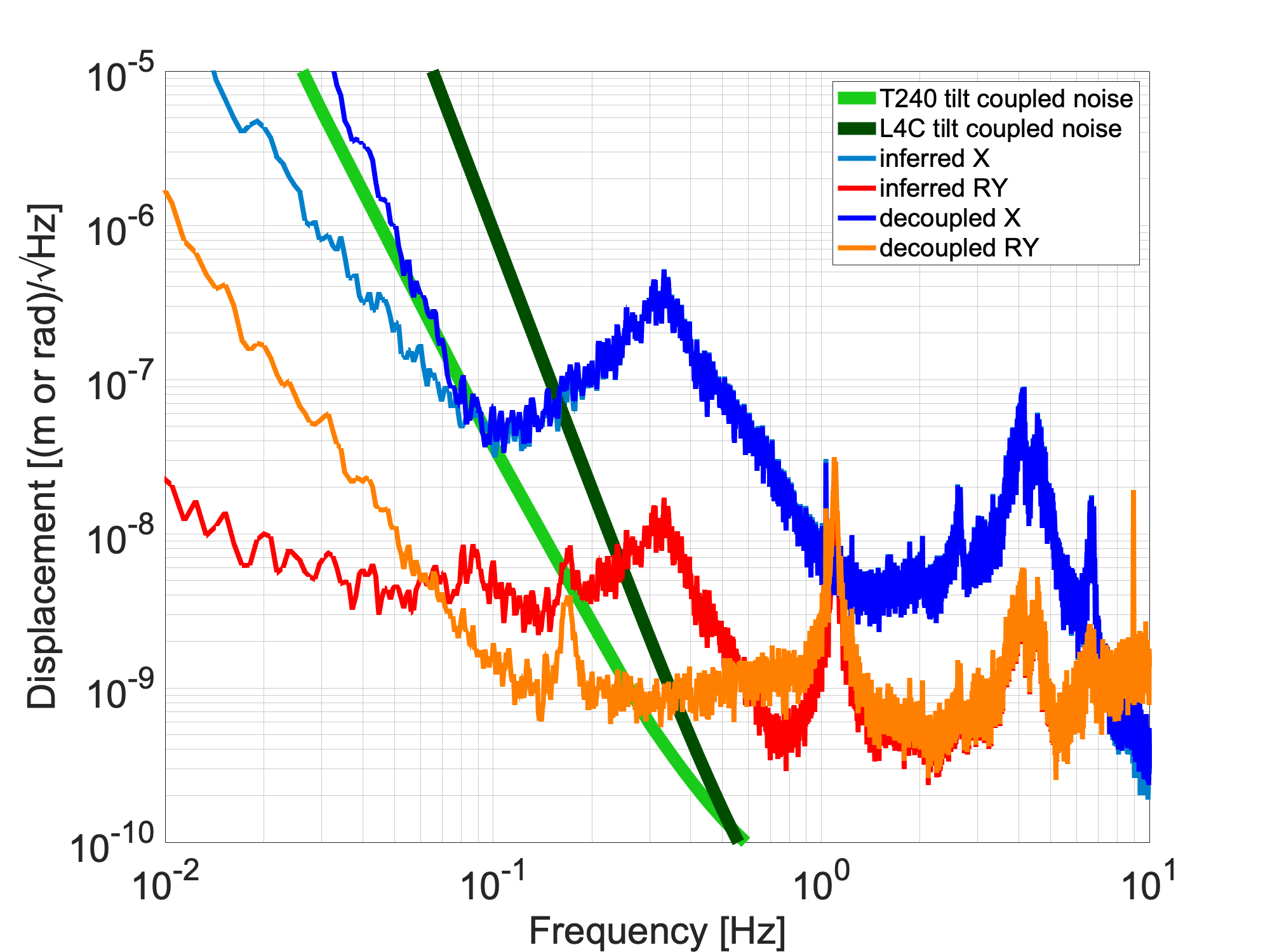}
        \caption{The figure shows the difference between the plant inversions of the translational and tilt modes with and without taking into account the X and RY cross-couplings. The orange curve shows the translation motion decoupled from the tilt mode such that the microseismic motion is removed. From the current sensitivity of the prototype, we see that the decoupled translational motion has the same noise floor as the tilt coupled T240 noise (green), and has better sensitivity than the L4C (dark green).}
        \label{fig:decoupling}
    \end{figure}

     In the laboratory conditions, apart from ground motion, the soft suspension is susceptible to sources of environmental noise, such as air pressure and density fluctuations. The fluctuations move the test mass and couple to the position readout of the mass. To minimise these effects, we placed the system inside a foam padded box that acted as a shield from the local air currents. The box improved the environmental noise by a factor of 30 at 10\,mHz. A further reduction by of factor 100 was achieved by back filling the box with foam pellets seen in Fig.\,\ref{fig:prototype}. Despite this reduction of environmental noises, our tilt measurement was still limited by air currents below 100\,mHz. We expect that environmental noises will disappear once we install the seismometer in vacuum.

    Apart from the environmental noise sources, our readout signals measured the platform motion since the interferometric readout noise and thermal noises were significantly smaller compared to the ground vibrations in our experiment. We compared the readouts of our seismometer with commercial L4C sensors~\cite{L4Cwebsite} as shown in Fig.~\ref{fig:spectra}. The L4Cs were co-located and co-aligned with the 6D seismometer to maximise coherence between the sensors. Fig.~\ref{fig:spectra} shows the comparison between measured signals before plant inversion. The spectra for the HoQI horizontal and L4C horizontal are comparable because the resonances for both devices are the same ($\approx$ 1\,Hz). The Q factor for our seismometer is larger than that of the L4C hence Fig.~\ref{fig:spectra} shows the more prominent 1\,Hz resonance in the HoQI spectra. From these measurements we find that the compact 6D noise floor in air is already better than that of the L4Cs below 100\,mHz.

    % \item \textbf{HoQI on suspension}\note{SAM section}

    % Demonstrated that HoQI sensors are capable of working on soft suspension systems. Imperative to keep RMS motion of TM small so that the system remains in the sensors working range. Linearisation can drastically improve the broadband noise and up-conversion of large motion in the spectra, see \href{https://lnx0.sr.bham.ac.uk:8081/Compact+6D/55}{elog entry}. Dr S. Cooper is investigating implementation of linearisation in real time using an ellipse fitting algorithm in CDS. \note{cite Sam's papers of chapter 3 of thesis}
    
    According to our estimation of the platform translational and tilt motion, the horizontal HoQI sensor measures tilt motion below 100\,mHz and translational motion at the other frequencies. The vertical HoQI sensor measures tilt and inertial mass motion due to environmental noise below the tilt resonance at 85\,mHz. Translational motion couples to the vertical tilt readout between the tilt and translational resonances of the suspension according to the transfer functions shown in Fig.~\ref{fig:transfer-functions}. At frequencies from 1.5\,Hz to 10\,Hz, the vertical sensor measures the platform tilt motion, and vertical platform motion couples to the readout at higher frequencies.
    
    We have diagonalised our readout signals according to Eq.~(\ref{eq:decoupling}) as shown in Fig.~\ref{fig:decoupling}. The tilt-decoupled horizontal motion $\hat{x}$ is larger than the inferred horizontal motion without tilt subtraction since our tilt measurement is limited by the environmental noises below 100\,mHz. However, the achieved sensitivity is comparable to the tilt-decoupled sensitivity of the Trillium T240 sensors~\cite{T240wesbite} that are currently installed in the aLIGO detectors. We calculated the tilt-decoupled power spectrum of the T240 sensor noise according to the equation
    \begin{equation}
        S_{\rm T240}^{x} = S_{\rm T240} \left( 1 + \frac{2 g^2}{\omega^4 \Delta L^2} \right),
    \end{equation}
    where $\Delta L = 1\rm{m}$ is the separation between two vertical T240 seismometers used for the measurement of the platform tilt motion, $S_{\rm T240}$ is the power spectrum of the T240 own noise, $g/\omega^2$ is the tilt-to-horizontal coupling coefficient, and a factor of 2 in the brackets accounts for two T240 seismometers utilised in the tilt measurement. 
    
    %The measured motion from the vertical HoQI shows the coupling between the translational and tilt motion. This coupling can be seen in Fig.~\ref{fig:transfer-functions}, where the gain is maximal between the pitch and translational resonance where the microseism is apparent. We see that in Fig.~\ref{fig:decoupling} the translational component of the vertical tilt measurement is removed via reduction of the microseismic motion. The decoupling is not yet ideal as the 1\,Hz resonance is not fully decoupled. Further investigation into the system identification to ideally model the transfer functions is required to improve the current decoupling. 
    
    %The decoupling shown makes it appear as if the noise when decoupling the signals (blue and orange) is larger than that with a regular plant inversion. However for comparison the tilt coupled noise floors for the L4C and T240 seismometers have been included (greens) to show what the real noise floor is when there is tilt motion which plagues the signal. The regular plant inversion of the translational motion (light blue) shows what the theoretical noise floor if there was purely translational motion, which is not the case for a real system.
    
    Fig.~\ref{fig:decoupling} also shows the decoupling of the horizontal platform motion from the tilt measurement at frequencies between 85\,mHz and 1\,Hz. Apart from the platform tilt motion, we can also see coupling from the other degrees of freedom in $\hat{\theta}$. The resonance at 167\,mHz corresponds to the RX eigen mode and the peak at 1.09\,Hz corresponds to the coupling of the Y motion. We conclude that our sensors were not ideally aligned to measure only X and RY degrees of freedom. We leave this coupling problem for further experimental investigations.
    
    %The measured tilt motion (orange) shows the coupling of the translational motion in to the tilt, when the decoupling is applied, the contribution is removed from the signal, therefore a purer tilt measurement is achieved. Placing the system on an active platform would cause the tilt motion to become more dominant in the translation signal. The decoupling would show that this tilt motion would then be removed from the translational signal as has been achieved for the tilt measurement.

\section{Fused silica seismometer}\label{sec:design}

From the lessons learned in Sec.~\ref{sec:prototype}, we propose to build the seismometer for GW detectors from fused silica and suspend the mass using a fused silica fibre, similar to the aLIGO test masses~\cite{Cumming2012}. The design is compact ($\approx 0.5$\,m scale) compared to the larger ($\approx 1.2$\,m) scale 6D seismometer~\cite{MowLowry2019} that the authors investigated in collaboration with VU, Amsterdam. The compact design allows us to (i) achieve low drift rates of the suspended mass as discussed in Sec.~\ref{sec:lf_drifts}, (ii) utilise commercially available fused silica masses, and (iii) remove the angular control noise from the aLIGO sensitivity band.

We choose fused silica for the design of the compact 6D seismometer due to its low thermal expansion coefficient and low mechanical dissipation. In this section, we consider two designs of the seismometer with the tilt eigenmode at 50\,mHz and 100\,mHz, find the own noise of the seismometer, derive the aLIGO platform noise, and estimate the angular motion of the aLIGO test masses. We also compare our results with the performance of the currently installed T240 seismometers.

    \begin{figure}[t]
        \centering
        \includegraphics[width = 0.9\columnwidth]{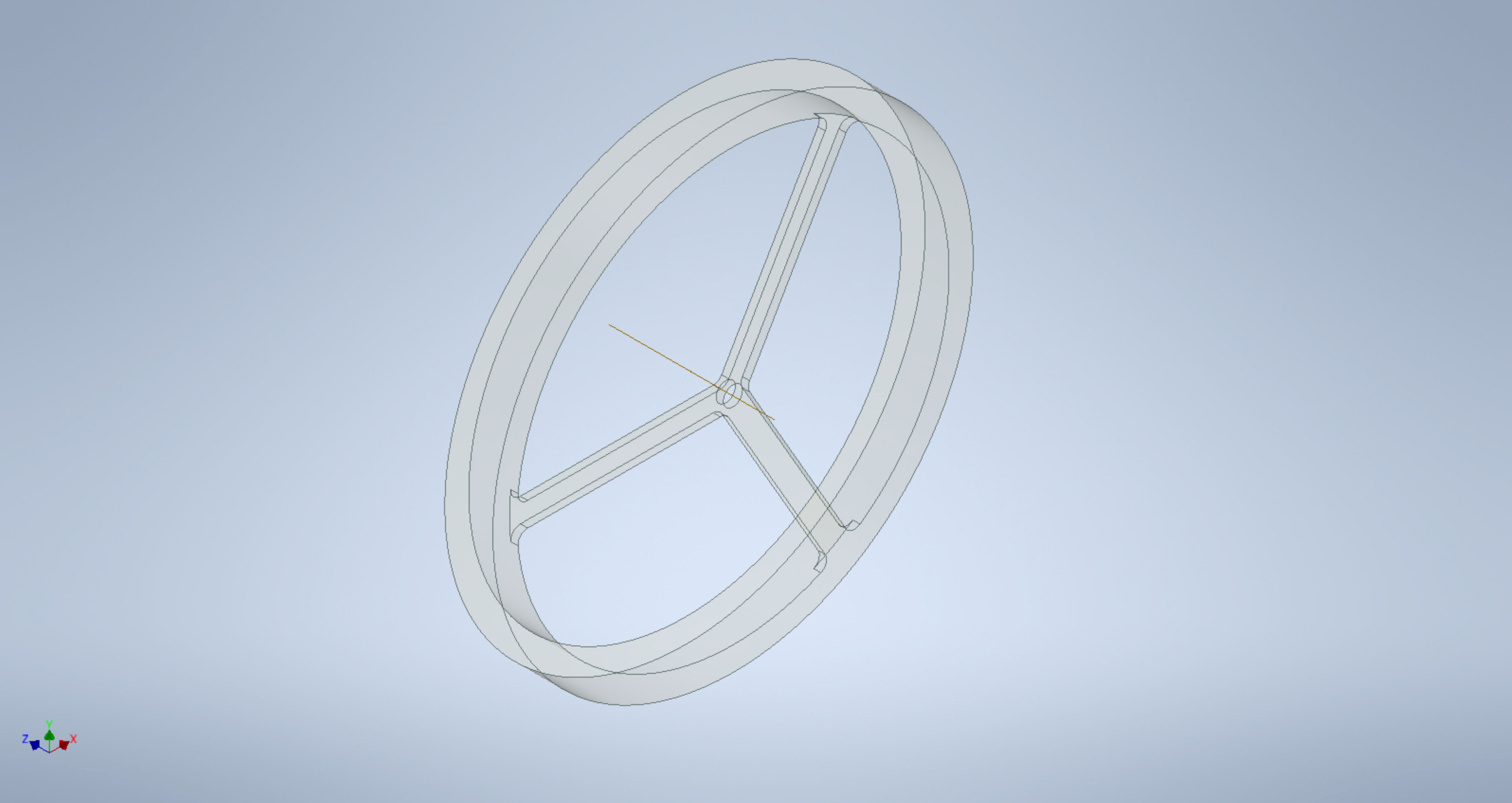}
        \caption{Model of the proposed fused silica compact 6D seismometer for the aLIGO active isolation platforms.}
        \label{fig:frame-mass-assy}
    \end{figure}
    
    \begin{table}[t]
        \caption{A list of parameters and nominal values}
        \begin{tabular}{ccc}
        \hline
        \hline
        Parameters & Description & Value\\
        \hline
        $m$ & Mass & 3.25\,kg \\
        \hline
        $R$ & Mass radius & 0.25\,m \\
        \hline
        $L$ & Wire length & 0.4\,m \\
        \hline
        $r$ & Fibre radius at bending points & 120\,$\mu$m \\
        \hline
        $f_{\rm X,Y}$ & Translational resonance & $0.8$\,Hz\\
        \hline
        $f_{\rm RX,RY}$ & Tilt resonance & 50\,mHz or 100\,mHz\\
        \hline
        \hline
        \end{tabular}
        \label{tab:parameters}
    \end{table}

Fig.\,\ref{fig:frame-mass-assy} shows the design of the fused silica suspension and Table~\ref{tab:parameters} summarises the key parameters of the design.
%Commercial companies, such as Heraeus \todo{cite Heraeus}, are now capable of machine blocks of fused silica up to $0.5\text{m}$ in diameter.
The customised commercially available fused silica mass~\cite{Heraeus} is circular in shape with the mass concentrated at the edges, and the three arms are designed to be as minimal as possible to reduce mass near the centre, therefore, maximising the moment of inertia for this configuration. The vertical and horizontal position of the centre of mass will be tuneable via small masses.

\subsection{Low frequency drifts}
\label{sec:lf_drifts}

    Ideally, the angular modes of the suspended inertial mass should be as low as possible to diagonalise the tilt and translational signals of the seismometer. However, soft suspensions tend to drift at low frequencies due to stress relaxations in the fibre~\cite{Levin_2012} and temperature gradients along the suspended mass. Strong drifts have the potential to move the system out of its interferometric sensing range. Experiments with torsion balances conducted in Washington~\cite{Adelberger1990}, Paris, and Birmingham~\cite{Quinn_2013}, and at the LISA facilities in Florida~\cite{Ciani2017} and Trento~\cite{Russano2016} show that the fused silica suspensions drift less than the tungsten ones and that softer RZ eigenmodes lead to larger drifts. 
    
    We propose to stiffen RZ by making the fibre cross-section non-uniform: its radius will be $r=120\,\mu$m near the fibre ends and $r=500\,\mu$m near the centre of the fibre. The RZ eigen mode is then given by the equation
    \begin{equation}
        f_{\rm RZ} = \frac{1}{2\pi}\sqrt{\frac{E_t J}{L_{\rm eff} I_{\rm RZ}}},
    \end{equation}
    where $E_t=30$\,GPa is the modulus of torsion of fused silica, $J = \pi r^4 / 2$ is the second moment of area along the axis of the cylindrical fibre, $L_{\rm eff}$ is the length of the fibre section with radius $r=120\,\mu$m. To stiffen RZ, the effective fibre length, $L_{\rm eff}$, must satisfy the following condition
    \begin{equation}
    \label{eq:l_eff}
        2 \Delta = 2 \sqrt{\frac{E I_a}{m g}} \ll L_{\rm eff} \ll L,
    \end{equation}
    where $\Delta = 600\,\mu$m is the bending length of the fibre, $L=40$\,cm is the fibre length, and a factor of 2 accounts for two bending locations of the fibre. We choose $L_{\rm eff} = 1$\,cm to satisfy Eq.~(\ref{eq:l_eff}) and get $f_{\rm RZ}=31$\,mHz. We expect the drift rate of less than $1\,\mu$m/week due to stress relaxation in the fibre.

    Apart from the fibre unwinding, we also consider thermal gradients across the inertial mass. The gradients make different parts of the mass expand non-uniformly and cause the suspension to lose its balance. We estimate the angular deviation in RX and RY according to the equation
    \begin{equation}
        \theta_{\rm rms} \approx \frac{g \alpha \Delta T_{\rm rms}}{2 R \omega_{\rm RX, RY}^2} = 1.6 \times  \frac{\Delta T_{\rm rms}}{20\, \rm mK} \left( \frac{50\, \rm mHz}{f_{\rm RX, RY}} \right)^2 \,\mu{\rm rad},
    \end{equation}
    where $\alpha=4 \times 10^{-7} \rm K^{-1}$ is the thermal expansion coefficient of fused silica~\cite{Heraeus}, and $\Delta T_{\rm rms}$ are thermal gradients along the inertial mass. We conclude that the thermal gradients are not significant for the proposed fused silica design.

\subsection{Projected motion of the aLIGO platforms}

\begin{figure}[t]
    \centering
    \includegraphics[width = \columnwidth]{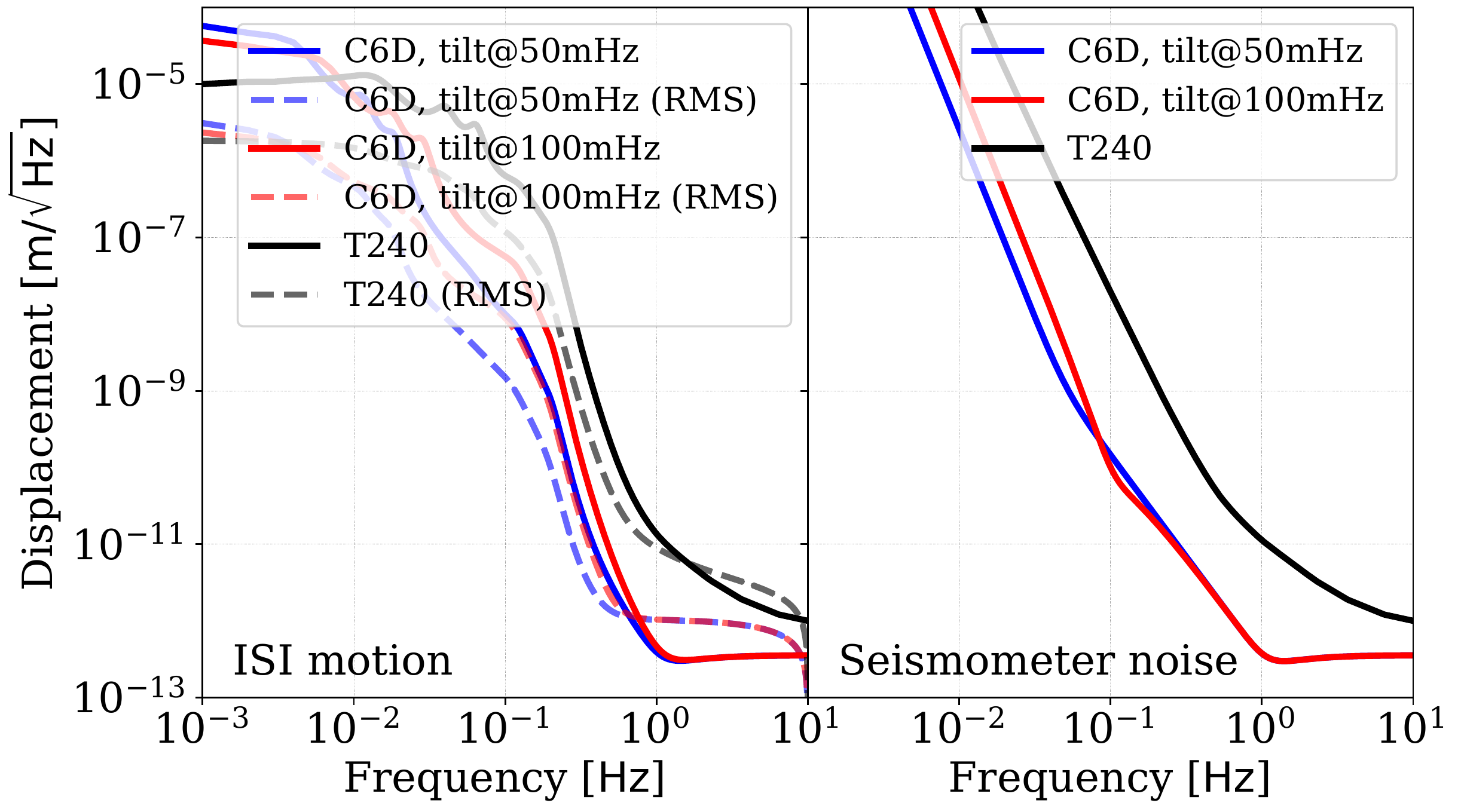}
    \caption{Estimated motion of the aLIGO platforms with the 6D seismometers (left). The right panel shows the comparison between the own noises of the 6D seismometers with the tilt eigenmode at 50\,mHz and 100\,mHz, and with the own noise of the T240 seismometers.}
    \label{fig:ISI}
\end{figure}

The sensitivity of the inertial mass position readout is limited by the own noise of the compact interferometer, thermal noise of the suspension, and the actuation noise. The readout noise will be the dominant noise of the seismometer. We set the requirement of
\begin{equation}
    S_{\rm ifo} = \left( 3 \times 10^{-13} \right)^2 + \left( \frac{10^{-13}}{f} \right)^2 \frac{\rm m^2}{\rm Hz}
\end{equation}
for the local readout of the mass position in the frequency band from 1\,mHz up to 10\,Hz. This level of sensitivity has been already achieved by the custom compact interferometers~\cite{Cooper_2018, Isleif_2019}.

Local readout noise couples to the estimated horizontal motion of the platform, $\hat{x}$. According to the dynamics of the system~\cite{MowLowry2019}, the readout noise coupling to $\hat{x}$ grows as $1/f^2$ below the longitudinal resonances, $f_{X,Y}$, and as $1/f^4$ below the tilt resonances, $f_{\rm RX, RY}$ as shown in Fig.~\ref{fig:ISI}. The $1/f^2$ and $1/f^4$ couplings are similar to the current noise scaling in the aLIGO detectors. For the T240 seismometers, the $1/f^2$ slope starts from 1\,Hz and the $1/f^4$ starts from the effective tilt resonance given by the equation
\begin{equation}
    f_{_{\rm RX, RY}}^{\rm T240} \approx \frac{1}{2\pi}\sqrt{\frac{2g}{\Delta L}} = 0.7\,{\rm Hz}.
\end{equation}

Our calculations of the suspension thermal noise of the inertial mass show that it becomes insignificant for the quality factors of the tilt mode $Q_t > 2 \times 10^4$. Since the aLIGO suspensions achieve loss angles of the suspension fibres below $10^{-6}$~\cite{Cumming2020}, the required quality factors can be achieved with the proposed fused silica suspension. We propose to damp the high quality resonances with magnet-coil actuators similar to the aLIGO suspensions~\cite{Strain2012}. The damping will allow us to reduce the quality factors of the suspensions down to $\sim 10$ and simplify the diagonalisation of the readout signals according to Eq.~(\ref{eq:decoupling}). Since we need to make the actuator noise insignificant, the actuator range will be $~10\,\mu$m and will not be able to correct any strong low-frequency drifts. However, as discussed in Sec.~\ref{sec:lf_drifts}, the fused silica design of the suspension will help avoid strong drifts.

The total noise of the seismometers, shown in Fig.~\ref{fig:ISI}, becomes larger than the ground motion below $\approx 30$\,mHz. In order to avoid large motions of the aLIGO platforms below 10\,mHz, aLIGO utilises a blended control scheme~\cite{Matichard_2015, Matichard2016}. The aLIGO platforms follow the signals from the inertial sensors only above the blending frequency. At lower frequencies, the aLIGO platforms follow the signals from the positions sensors that measure the relative motion between the aLIGO platforms and the ground. 
%the signals are filtered with a so-called blending filter, \textit{i.e.} a high-pass filter for the seismometer signal and a low-pass filter for ground motion signal. The whole sensing scheme is caller super sensor.
A lower blending frequency provides better seismic isolation above 100\,mHz but couples more inertial sensor noise to lower frequencies. In the aLIGO detectors, where T240 seismometers are currently deployed, the blending frequency is set at 45\,mHz.  The proposed 6D seismometer has the potential to lower the blending frequency by a factor of $2-3$ without increasing the RMS motion of the platform as shown in Fig.~\ref{fig:ISI}. In this study, we find that a 6D seismometer with a tilt mode at 50\,mHz (100\,mHz) can reduce the blending frequency down to 14\,mHz (22\,mHz).
%lower blending frequency without being saturated by the inertia sensor noise at high frequencies. As discussed above, the target for the pitch mode is 50\,mHz. In that case, we choose blending frequency at 13\,mHz, which give similar rms motion, 3 micrometers, to the case of Advanced LIGO. The residual ISI platform motion and the seismoter noise as a sum of the inertia sensing noise and the thermal noise are shown in Fig.~\ref{fig:ISI} 

\subsection{Projected motion of the aLIGO test masses}

\begin{figure}[t]
    \centering
    \includegraphics[width = \columnwidth]{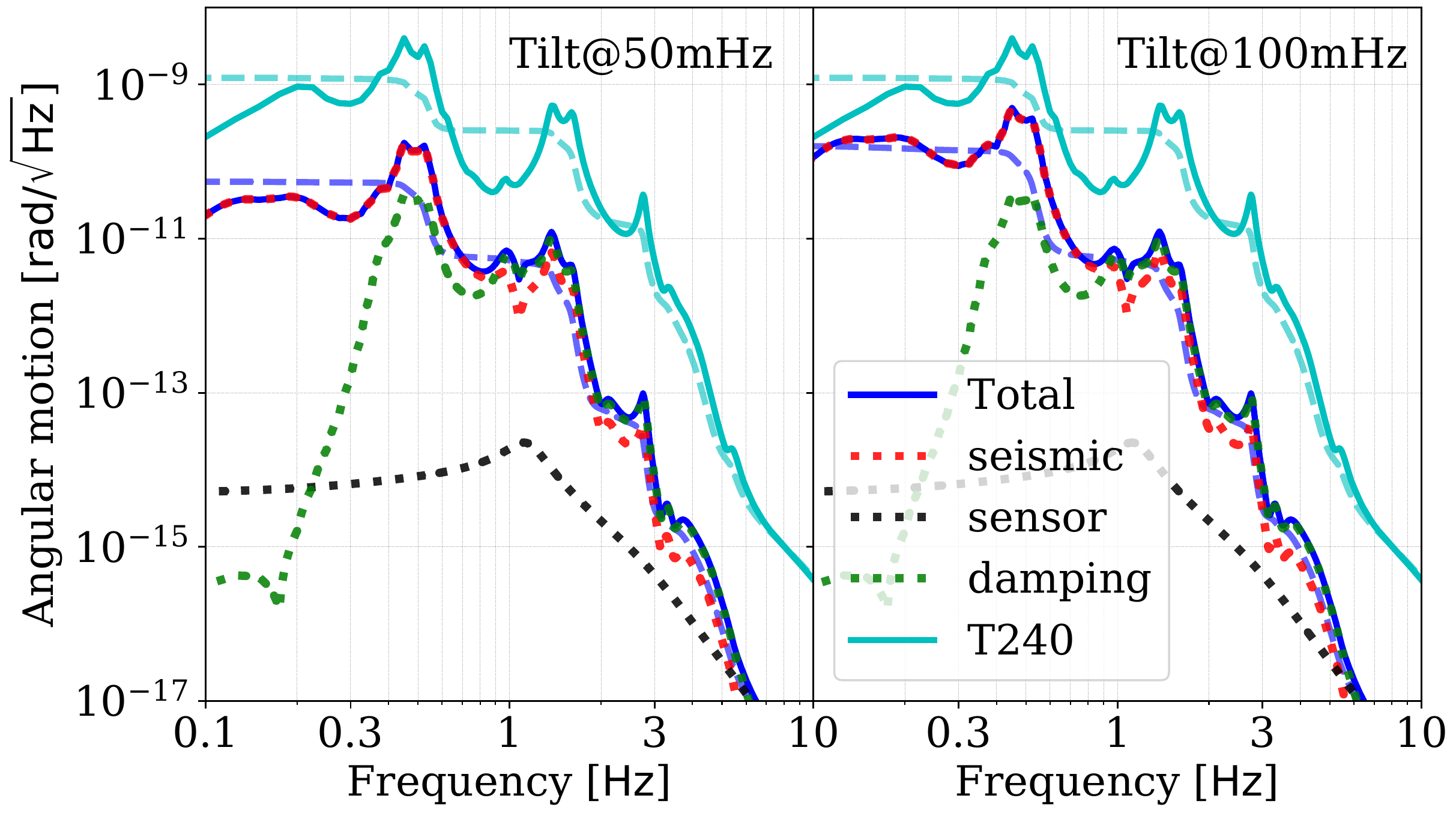}
    \caption{Comparison of the estimated pitch motion of the aLIGO test masses when the aLIGO platforms are stabilised with the proposed 6D seismometers with a tilt eigenmode of 50\,mHz (left panel), 100\,mHz (right panel) and the T240 seismometers.}
    \label{fig:MirrorAngular}
\end{figure}

The key noise source of the aLIGO detectors at low frequencies (below 30\,Hz) is the angular control noise. The motion of the aLIGO platforms causes angular motion of the test masses. The motion is then actively stabilised by wavefront sensors~\cite{Mavalvala_ASC_1998, Barsotti_ASC_2010}. The control loops keep the relative motion of the aLIGO test masses below $\sim 1$\,nrad, but couple the wavefront sensor sensing noise to the aLIGO GW channel~\cite{Dooley_ASC_2013}. The bandwidth of the angular control loops is currently set to be 3\,Hz~\cite{Martynov_Noise_2016}. Since the 6D seismometer has the potential to improve the motion of the aLIGO platforms as shown in Fig.~\ref{fig:ISI}, the angular controls bandwidth can be reduced by a factor of 3 and remove coupling of the aLIGO angular controls to its sensitivity band above 10\,Hz~\cite{Yu2018}.

In this section, we estimate the angular motion of the aLIGO test masses with the 6D seismometer. The dominant angular motion of the aLIGO test masses comes from the longitudinal motion of the aLIGO platforms~\cite{Yu2018}. We propagate the platform motion through the transfer functions of the aLIGO quadruple suspensions and compute the motion of the test masses. We find that we can achieve residual relative motion of the aLIGO test masses below 1\,nrad with a control bandwidth of 1\,Hz and a typical wavefront sensor noise of $5\times 10^{-15}\, {\rm rad}/\sqrt{\rm Hz}$ when the tilt mode of the seismometer inertial mass is either 50\,mHz or 100\,mHz. The results of mirror angular motion are shown in Fig.~\ref{fig:MirrorAngular}. The damping noise comes from the local sensors that damp the high quality factors of the aLIGO suspensions. In computing the damping noise for Fig.~\ref{fig:MirrorAngular}, we assume that the aLIGO suspensions are equipped with the same interferometeric positions sensors as the 6D seismometers.

We find that both configurations of the 6D seismometer from Table~\ref{tab:parameters} have the potential to solve the problem of the aLIGO angular control coupling to the GW channel below 20\,Hz. The 100\,mHz configuration is easier to implement compared to the 50\,mHz one since center-of-mass tuning $\Delta z$ can be less precise. However, the 50\,mHz version provides significantly better inertial isolation in the earthquake band between 30\,mHz and 300\,mHz, and has the potential to keep the aLIGO detectors operational during small and medium earthquakes (local RMS velocity of the ground below $\sim 1\mu \rm m/s$) and further improve the aLIGO duty cycle.

\section{Conclusion}

Novel inertial sensors can enhance the sensitivity of the aLIGO detectors below 30\,Hz, improve their duty cycle, and open a pathway towards increasing the beam size in the aLIGO arms. In this paper, we presented experimental results from our studies of the compact 6D seismometer with an interferometric readout. We conclude that the seismometer has the potential to improve aLIGO's seismic isolation. Measurement of the horizontal and tilt modes of the metallic prototype resulted in validation of the decoupling capabilities of the readout scheme though decoupling around the horizontal eigen mode requires further investigations.
%In this paper we demonstrate a proof of concept of a compact six degree of freedom seismometer which could be implemented at the LIGO sites. A prototype was constructed in order to understand the mechanics, dynamics, and complexities of such a system to provide insight for a final vacuum compatible LIGO approved design. 
We found that lower tilt eigenmodes of the suspended mass make the balancing procedure complicated due to plastic deformations in the wires.

Using the lessons learned from the metal prototype, we proposed and simulated the performance of a fused silica compact 6D seismometer with the tilt modes at 50\,mHz and 100\,mHz. We found that both configurations should have low angular drift rates due to the mechanical and thermal properties of fused silica. The seismometers are capable of improving the current aLIGO low frequency noise but the 50\,mHz design also provides strong suppression of the the ground motion in the earthquake band. Experimental investigations of the compact fused silica 6D seismometer will be our next step.

%The investigation of the translational and pitch mode resulted in validation of the decoupling capabilities of the readout scheme shown in Fig.~\ref{fig:decoupling}. This enables improved sensing of the proof mass down to the pitch eigenmode. Improving the platform motion using a device such as compact 6D would reduce the motion of the LIGO test masses, which would decrease the strain on the angular control loops \cite{Yu2018}. From Fig.~\ref{fig:ISI} and \ref{fig:MirrorAngular} we show the predicted displacement and angular motion using compact 6D is superior to the current LIGO configurations using commercial T240 seismometers.

% This proof of concept provides confidence in constructing a LIGO approved final design which could be implemented in future upgrades at the Livingston and Hanford sites.

\section*{Acknowledgements}
We thank Hang Yu, members of the LIGO SWG group and our collaborators on 6D seismometers from VU, Amsterdam for useful discussions.
The authors acknowledge the support of the Institute for Gravitational Wave Astronomy at the University of Birmingham, STFC 2018 Equipment Call ST/S002154/1, STFC 'Astrophysics at the University of Birmingham' grant ST/S000305/1. A.S.U. and J.S. are supported by STFC studentships 2117289 and 2116965.  H. M. is supported by UK STFC Ernest Rutherford Fellowship ST/M005844/11. 

\section*{Bibliography}
\bibliographystyle{unsrt}
\bibliography{main.bib}

\end{document}